\documentclass[11pt,twoside]{article}
\usepackage{./asp2014}

\aspSuppressVolSlug
\resetcounters

\begin{document}

\title{MaNGA: Mapping Nearby Galaxies at Apache Point Observatory}
\author{Anne-Marie Weijmans$^1$ on behalf of the MaNGA team
\affil{$^1$School of Physics and Astronomy, University of St Andrews,
  North Haugh, St Andrews, KY16 9SS, UK; \email{amw23@st-andrews.ac.uk}}}

\paperauthor{Sample~Author1}{Author1Email@email.edu}{ORCID_Or_Blank}{Author1 Institution}{Author1 Department}{City}{State/Province}{Postal Code}{Country}

\begin{abstract}
MaNGA (Mapping Nearby Galaxies at APO) is a galaxy integral-field spectroscopic survey within the fourth generation Sloan
Digital Sky Survey (SDSS-IV). It will be mapping the composition and
kinematics of gas and stars in 10,000 nearby galaxies, using 17 differently sized fiber
bundles. MaNGA's goal is to provide new insights in galaxy formation
and evolution, and to deliver a local benchmark for current and future
high-redshift studies. 
\end{abstract}

\section{Introduction}

Multi-object spectroscopy plays a large role in current research on
galaxy formation and evolution. The Sloan Digital Sky Survey (SDSS) has
provided spectra for a million local galaxies
\citep{york2000,strauss2002}, enabling us to study large samples of
galaxies, and catch galaxies in rare and short-lived phases of galaxy
evolution. By measuring the integrated spectroscopic properties of
galaxy centres, large spectroscopic surveys have shown how galaxy
properties such as morphology, age, metallicity and star formation
rates vary with mass, environment and redshift.

In addition, integral-field or spatially resolved spectroscopy has
pushed our understanding of galaxies further. Not limited to single
fiber spectroscopy covering only the central region of the galaxy,
integral-field surveys provide us with resolved spectroscopic
properties over a large galaxy field-of-view. The SAURON
survey \citep{dezeeuw2002} mapped the gas, stellar populations and kinematics of 72
local ($z \leq 0.01$) early-type galaxies, and with the ATLAS3D
survey \citep{cappellari2011} of 260 early-type galaxies, introduced a
new kinematical classification scheme
\citep{emsellem2007,emsellem2011}. The currently on-going CALIFA
survey \citep{sanchez2012} includes both early- and late-type
galaxies, with a total sample size of 600 galaxies.

To make larger ($\sim$thousands of galaxies) integral-field surveys
possible, multi-object integral-field spectroscopy is the way forward.
The SAMI survey \citep{croom2012}
uses a multiplexed fiber integral-field unit (IFU) to observe a total
of 3400 galaxies \citep[see also][]{bryant2015}. With MaNGA \citep{bundy2015} we will be measuring resolved stellar and gas
properties for a total of 10,000 galaxies. Apart from collecting
millions of spectra to study galaxy evolution in local galaxies, MaNGA
will also provide a much-needed local benchmark for current and future
high-redshift studies, such as with MUSE \citep{bacon2010} and MIRI
  \citep{wright2004}. MaNGA is part of
SDSS-IV\footnote{www.sdss.org} and will run from 2014 to 2020, with
regularly scheduled public data releases. In this proceeding we
discuss MaNGA's science goals (\S2), instrument (\S3) and survey
design (\S4). We also highlight some of the early science papers based
on a MaNGA prototype observing run in \S5.

\section{MaNGA science goals}

MaNGA aims to map the kinematics and composition of the stars and gas in 10,000 nearby galaxies out to at least
1.5 half-light or effective radius $(R_e)$, to study the birth, death
and life of galaxies. In particular, MaNGA  will search for answers to
key questions such as: How are galaxy discs growing at the present
day, and where does the required gas for this growth come from? What
different processes are at play in the present growth of bulges and
elliptical galaxies? How does quenching of star formation occur, and
what role do internal and external processes play, including the
environment? How are mass and angular momentum distributed in the
different galaxy components, and how were these components affected
during galaxy assembly processes over time? These science goals have
been outlined in more detail in \citet{bundy2015}.

The MaNGA instrument, observing strategy and galaxy sample have been
carefully designed to provide answers to these questions. The sample
size of 10,000 galaxies has also been carefully considered: we want to
study galaxies as a function of stellar mass, environment and star
formation rate. Binning our galaxy sample in these three parameters,
we want a minimum of 50 galaxies in each bin, for $5\sigma$ detections
(the significance of a detected difference between two bins being
given as $\sqrt{n/2}$, with $n$ the number of galaxies in each
bin). If we require 6 bins in each of our three parameters, we
arrive at a required sample size of $50 \times 6^3 \sim$10,000 galaxies.

\section{Instrument design}

The MaNGA instrument is a multi-object integral-field unit
spectograph, consisting of 17 individual fiber IFUs. The fibers
within each IFU are hexagonally packed, and the IFUs range in
size from 12 arcsec diameter (19 fibers) to 32 arcsec diameter (127
fibers). These science IFUs are complemented with 12
mini-bundles consisting of just 7 fibers for the observations of
standard stars for flux calibration, and 92 sky fibers associated with
the science IFUs, for sky background
measurements. Fig.~\ref{fig:bundles} visualizes the different IFU
sizes. Individual fibers have an outer diameter of 150 $\mu$m and core
diameter of 120 $\mu$m, which translates to a diameter on sky of 2
arcsec. Deviations from a true hexagonal packing of individual fibers
within the IFU have an RMS value of 3 $\mu$m, or 2 per cent of the outer
diameter, and the filling factor is about 56 per cent. The IFUs
and their sky fibers are integrated into plug-plate cartridges, that
allow for a 3$^\circ$ field-of-view.  Table~\ref{tab:fibers} summarizes the MaNGA fiber
hardware for each cartridge, with a total fiber budget of 1423 per
cartridge, and 6 cartridges in use at Apache Point Observatory. More
details on the design, production and performance of the IFUs
are given in \citet{drory2015}.

\articlefigure{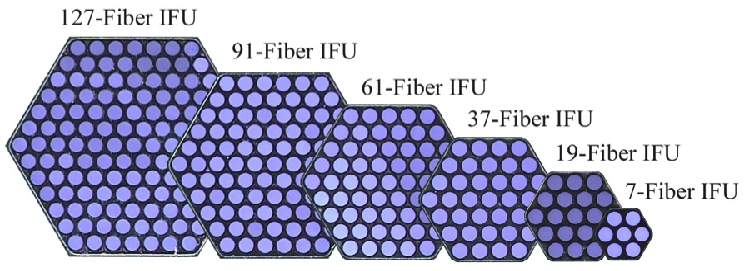}{fig:bundles}{Photographs of MaNGA IFUs, ranging from the largest IFU (127 fibers, left) to the
  mini-bundle used for standard star observations (7 fibers,
  right). Individual fibers deviate with an RMS of only 3 $\mu$m from an
  ideal hexagonal packing. Image taken from Drory et al. (2015).}

\begin{table}[ht]
\caption{Overview of the MaNGA instrument}
\smallskip
\begin{center}
{\small
\begin{tabular}{ccccc}
\noalign{\smallskip}
Fiber IFU & IFUs per cart & Sky fibers per IFU & Size on-sky (diameter) &
Comment \\
& & & (arcsec) &  \\
\noalign{\smallskip}
\tableline
\noalign{\smallskip}
127 &  5 & 8  & 32 &\\
91   &  2 &  6 & 27 & \\
61   &  4 &  4 &  22 &\\
37   &  4 &  2 & 17 & \\
19  &  2 &  2 & 12 & \\
7     &  12  & 1  & 7 & flux calibration mini-bundle \\
\noalign{\smallskip}
\tableline
\end{tabular}
}
\end{center}
\label{tab:fibers}
\end{table}

The MaNGA fibers are fed into the BOSS spectrographs,
which consist of an identical pair of dual beam spectrographs
\citep{smee2013}. With these spectrographs, MaNGA has a wavelength
coverage of 3600 - 10,300 \AA, with a spectral resolution varying from $R \sim
1400$ at 4000 \AA\ to $R \sim 2600$ around 9000 \AA.

\section{Survey design}

\subsection{Galaxy sample}

The MaNGA sample is constructed to have a flat distribution in stellar
mass, and to have a uniform radial coverage for all
galaxies. The fiber IFU distribution described above is optimized
for these requirements, and a detailed discussion will be presented in
Wake et al. (in prep). All MaNGA galaxies are selected from the SDSS
main spectoscopic sample, using an extension of the NASA-Sloan Atlas \citep{blanton2011}.

We defined a Primary and a Secondary sample, with the Primary sample
reaching a radial coverage of at least 1.5 $R_e$ (for at least 80 per
cent of the galaxies in this sample), and the Secondary sample reaching at least 2.5
$R_e$ (again for at least 80 per cent of the galaxies in this
sample), to study
emission line behaviour at larger radius. The Primary sample is
augmented with a colour-enhanced selection of galaxies, to increase
the number of high-mass blue and low-mass red galaxies, creating a
more balanced colour distribution for each stellar mass bin. The
colour-enhanced sample also includes green valley galaxies, tracing
rare or short-lived phases of galaxy evolution. This augmented sample
is our Primary+ sample. The median redshift of this sample is
$\langle z \rangle = 0.03$, leading to a spatial resolution of 1.3 - 4.5
kpc. The Secondary sample resides at slightly higher redshifts ($\langle z
\rangle = 0.045$) and has therefore lower spatial resolution (2.2 -
5.1 kpc). See \citet{bundy2015} for more details on these two
samples. Intially, the ratio between Primary+ and Secondary sample has
been set as 2/3 versus 1/3, but this ratio can be adapted over the
survey's lifetime.

\subsection{Observing strategy}

The observing strategy for MaNGA is described in detail in
\citet{law2015}. On average, the exposure time for a MaNGA plate to
reach a target signal-to-noise S/N of 5 \AA$^{-1}$ fiber$^{-1}$ is 3
hours, broken up in dithered sets of single exposures. We use a
three-point dither pattern, with three 15-minute exposures producing a
complete set. Dithering not only compensates for light losses between
individual fibers in the IFUs, but also improves the regularity of the
PSF (point-spread function) over the total area of the IFU, resulting
in an improved image reconstruction in the final datacube. This is
more quantatively illustrated in Fig.~\ref{fig:law}.

To ensure homogeneous quality of exposures within one set, we require
that the (S/N)$^2$ values in individual exposures within a set are
within a factor of two of each other, and that the difference in
atmospheric seeing within one set is less than 0.8 arcsec FWHM. The
average seeing within one set must be less than 2.0 arcsec FWHM to
ensure good spatial resolution. All exposures within one set must be
taken within one hour of each other, to minimize irregularities caused
by atmospheric refraction. \citet{law2015} has an extensive discussion
on these observing requirements, and shows detailed simulations
supporting our observing strategy decision. 

\articlefigure{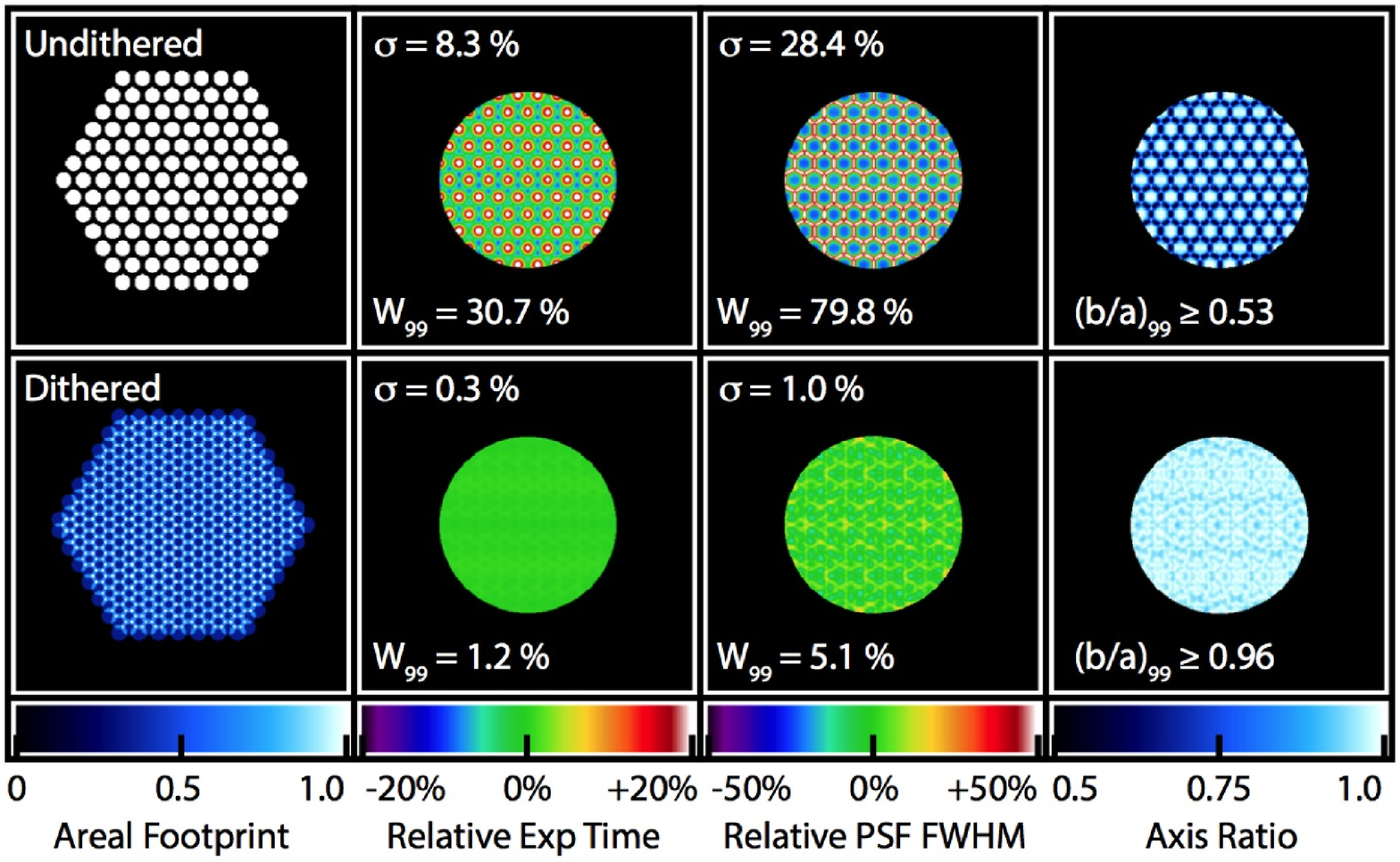}{fig:law}{The effect of dithering on PSF
  regularity and shape, illustrated with simulations. Top row: single
  exposure with 127-fiber IFU. Bottom row: dithered set of
  exposures with 127-fiber IFU. The first (most-left) column shows
  the footprint of the single fibers within the bundle. The second
  column illustrates the variations (in percentages) around the median exposure time
  within the bundle. The third column shows in percentages the
  deviations from the median FWHM of the PSF, while the fourth and
  most right colum shows the axis ratio $(b/a)$ of the PSF. The
  numbers in the top corner of columns 2 and 3 indicate the
  RMS deviations $\sigma$, while the numbers in the lower
  corner indicate the 3-$\sigma$ interval that contains 99 per cent
  of all values $W_{99}$. Dithering
  significantly improves the regularity of the PSF over the area of
  the IFU. Figure taken from Law et al. (2015).}

\section{First results: P-MaNGA}
To explore the instrument design and survey strategy for MaNGA we
developed a proto-type instrument (P-MaNGA), for test observations at Apache
Point Observatory in December 2012 and January 2013. Observing time
for this test run was generously donated by the SDSS-III
collaboration \citep{eisenstein2011}. P-MaNGA used only one
of the two available BOSS spectrographs, and consisted of eight IFUs (five
19-fiber IFUs, one 61-fiber IFU and two 127-fiber IFUs). The main
difference between P-MaNGA and MaNGA lies in the calibration fibers:
P-MaNGA had only 60 sky fibers distributed over two dedicated sky
V-groove blocks, and therefore not associated with the IFUs as in the
MaNGA design. P-MaNGA also did not have mini-bundles for standard star
observations, but instead used single fibers with larger diameters
(3.0 and 5.0 arcsec) to account for aperture losses (see Yan et al. in
prep). A detailed
description of the P-MaNGA instrument and observing run can be found in
\citet{bundy2015}.

Apart from several calibration frames, we also observed three galaxy
plates with P-MaNGA under various conditions, including high
airmass, to explore our survey limits. We observed 18 galaxies in total. These data were published in three P-MaNGA early science
papers, to explore the scientific potential of the larger MaNGA data
set. We summarize their results below, but refer to the papers
themselves for more detail.

\subsection{Emission line properties}

\citet{belfiore2015} presented emission line measurements for the
galaxies observed with P-MaNGA. They constructed spatially resolved
BPT \citep{baldwin1981} diagrams to characterize the ionisation
mechanisms responsible for the observed ratios in strong optical
emission lines. They also included a comparison with stellar popuation
indices such as the 4000 \AA\ break $D_n(4000)$ and the equivalent
width (EW) of H$\delta_A$ to gain insight into the connection between
the stellar population and ionisation parameters. Three of the
fourteen galaxies in their study harboured extended LINER-like
emission in their centres, as well as high $D_n(4000)$ and low
EW(H$\delta_A$) values, hinting at low star formation activity and an
old and metal-rich stellar population. These galaxies are consistent
with the so-called 'inside-out' growth scenario. Furthermore,
\citet{belfiore2015} found star formation activity in the extended discs of some
galaxies that are dominated by Seyfert- and LINER-like emission in
their centre, highlighting the advantage that IFU observations have
over single fiber data.

\subsection{Star formation gradients}

\citet{li2015} provided a study of the star formation history
properties of the P-MaNGA galaxies. They divided their galaxy sample
in centrally quiescent and centrally star forming galaxies, based on
$D_n(4000)$ value of their central spaxel, and mapped radial
profiles in $D_n(4000)$, EW(H$\delta_A$) and EW(H$\alpha$). The
centrally quiescent galaxies show no or only weak gradients in these
indices. In contrast, the centrally star-forming galaxies have
negative $D_n(4000)$ gradients, and positive EW(H$\delta_A$) and
EW(H$\alpha$) gradients, indicating a quenching of star formation that
starts in the central regions of the galaxies, and subsequently
propagates outwards (see Fig.~\ref{fig:li}).

\articlefigurethree{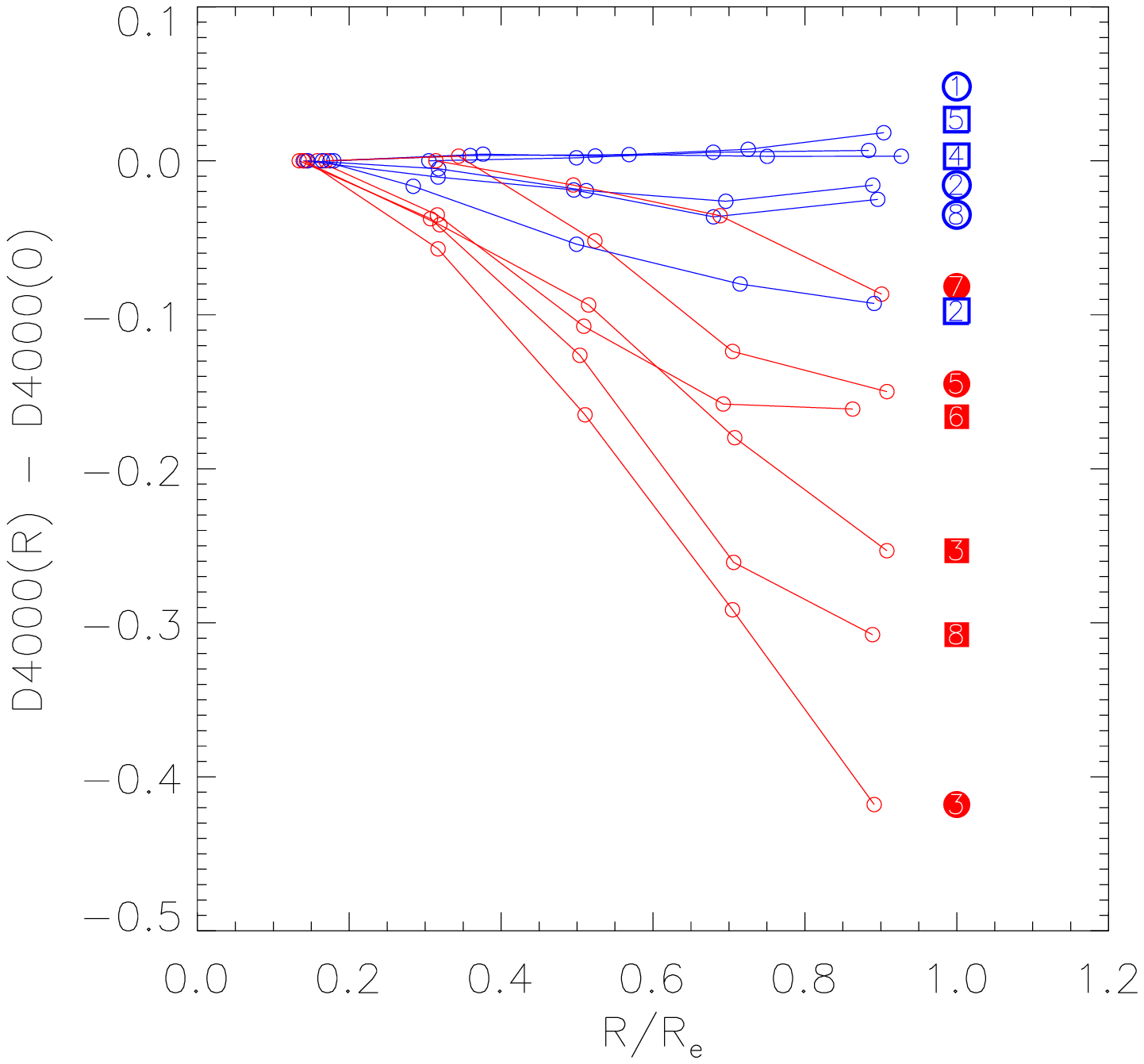}{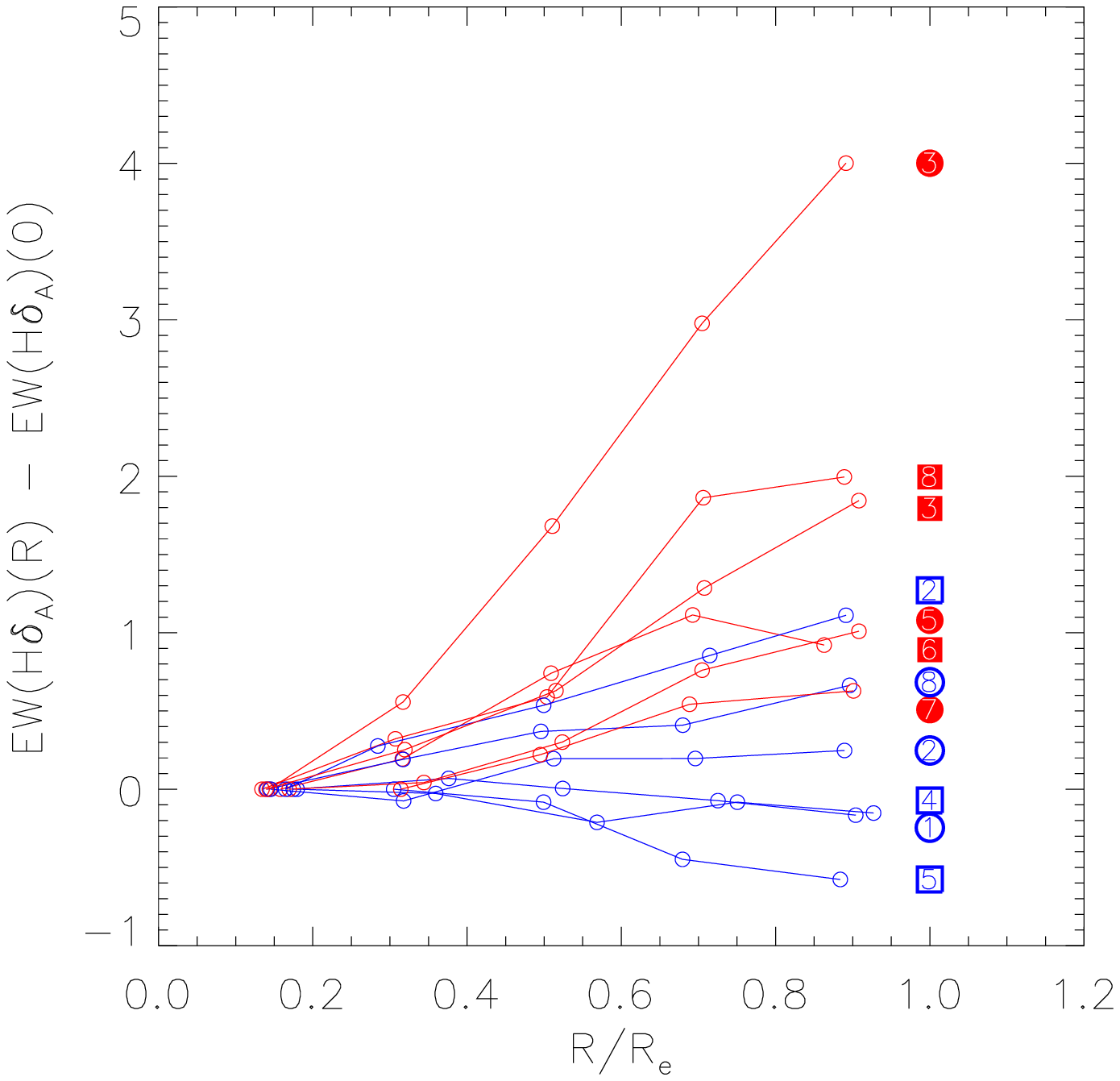}{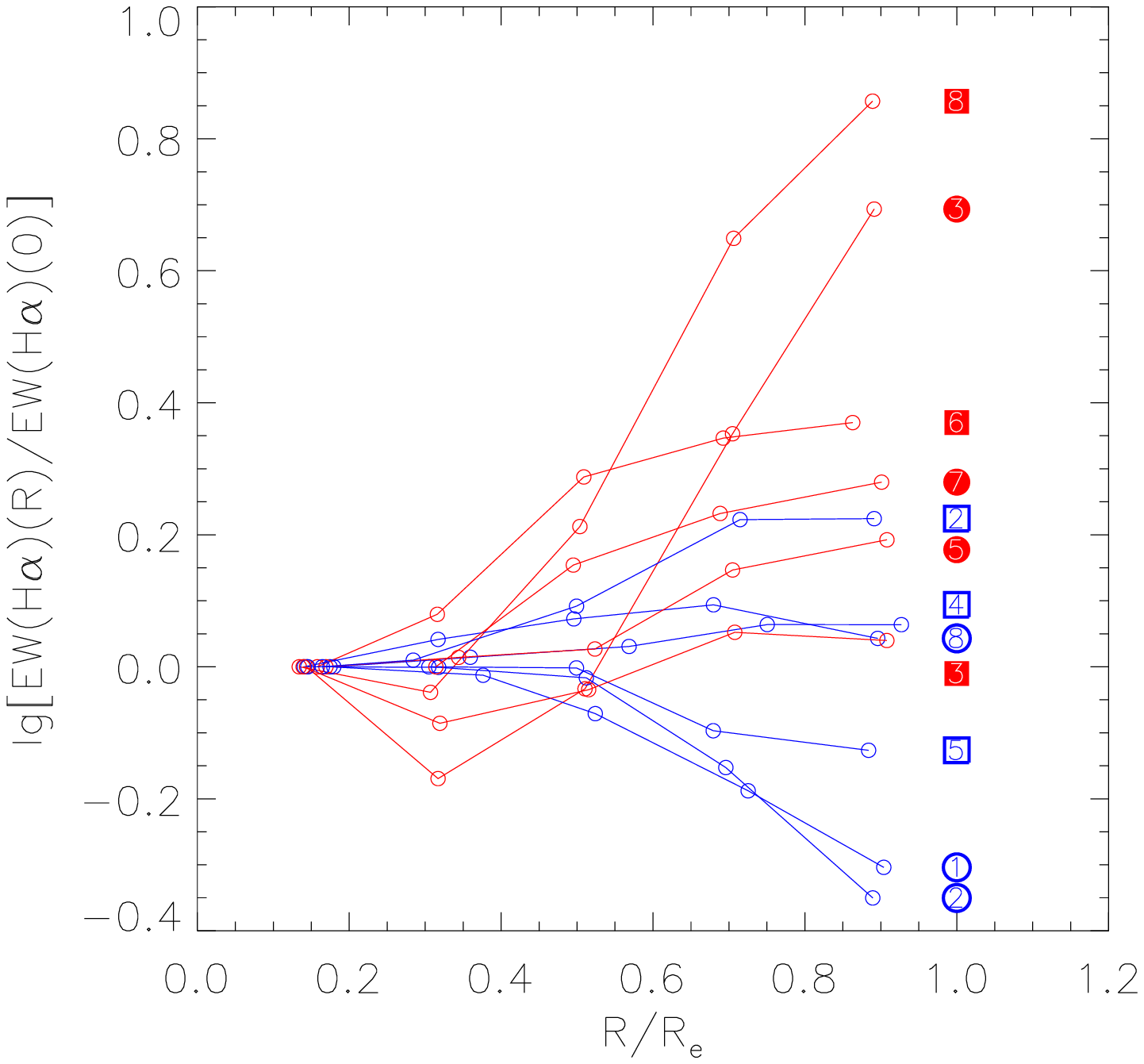}{fig:li}{From left to right: Radial profiles of
  $D_n4000$, EW(H$\delta_A$), and EW(H$\alpha$), as measured in the
  P-MaNGA galaxy sample. The
  profiles are normalised to the first radial bin. The centrally
  star forming galaxies (blue) show mostly flat profiles, while the
  centrally quiescent galaxies (red) have increasing $D_n(4000)$ and
  increasing EW of H$\delta_A$ and H$\alpha$. Figure taken from Li et al. (2015).}

\subsection{Stellar populations and dust}

\citet{wilkinson2015} produced detailed maps of stellar age,
metallicity and mass, as well as dust extinction, for the galaxies in
the P-MaNGA sample. They used their new full spectral fitting code
FIREFLY (Wilkinson \& Maraston, in prep) in combination with
high-resolution stellar population models by
\citet{maraston2011}. They recovered the dust displacement in stellar
population property maps, thereby breaking the degeneracies between
dust, age and metallicity. In addition to the maps,
\citet{wilkinson2015} derived radial gradients. They
found negative metallicity gradients and flat age gradients for
spheroidal galaxies, while in late-type galaxies the metallicity
gradient is flat and the age gradient is negative, consistent with
current literature
\citep[e.g.][]{yoachim2008,kuntschner2010}. \citet{wilkinson2015} also
provided an analysis of the effect of beam smearing on the derived maps
and gradients, and found that this effect is larger for the
smaller IFUs and more strongly affects light-weighted properties than
mass-weighted properties. The effects of beam smearing on
the different MaNGA IFU bundle sizes will be studied in more detail
for the full MaNGA sample.

\section{Conclusion}
We presented an overview of the MaNGA survey, which as part of SDSS-IV will map 10,000
galaxies with multi-object integral-field spectroscopy over its 6 year
survey lifetime. The MaNGA instrument, galaxy sample and observing
strategies have been optimized to fullfill the science goals as
described in this proceeding and in \citet{bundy2015}, and to make use
of already existing SDSS infrastructure. A successful observing run
with the P-MaNGA prototype instrument led to an improved
instrument and survey design, as well as
three early science publications. All MaNGA data will be
publicly available, and data releases will be
announced on the SDSS website.

\acknowledgements AW acknowledges support of a Leverhulme Trust Early
Career Fellowship. Funding for the Sloan Digital Sky Survey IV has
been provided by the Alfred P. Sloan Foundation and the Participating
Institutions. SDSS-IV acknowledges support and resources from the
Center for High-Performance Computing at the University of Utah. The
SDSS web site is www.sdss.org. SDSS-IV is managed by the Astrophysical
Research Consortium for the Participating Institutions of the SDSS
Collaboration including the Brazilian Participation Group, the
Carnegie Institution for Science, Carnegie Mellon University, the
Chilean Participation Group, Harvard-Smithsonian Center for
Astrophysics, Instituto de Astrofísica de Canarias, The Johns Hopkins
University, Kavli Institute for the Physics and Mathematics of the
Universe (IPMU) / University of Tokyo, Lawrence Berkeley National
Laboratory, Leibniz Institut für Astrophysik Potsdam (AIP),
Max-Planck-Institut für Astrophysik (MPA Garching),
Max-Planck-Institut für Extraterrestrische Physik (MPE),
Max-Planck-Institut für Astronomie (MPIA Heidelberg), National
Astronomical Observatory of China, New Mexico State University, New
York University, The Ohio State University, Pennsylvania State
University, Shanghai Astronomical Observatory, United Kingdom
Participation Group, Universidad Nacional Autónoma de México,
University of Arizona, University of Colorado Boulder, University of
Portsmouth, University of Utah, University of Washington, University
of Wisconsin, Vanderbilt University, and Yale University.

\end{document}